\documentclass[prd,showpacs,amsmath,amssymb]{revtex4}
\usepackage{amsmath}
\usepackage{amsfonts}
\usepackage{mathrsfs}
\usepackage{pstricks}
\usepackage{color}
\usepackage{graphicx}
\usepackage{slashed}
\usepackage{amssymb}
\usepackage{setspace}

\fontsize{18pt}{20pt}

\newcommand{\be}{\begin{equation}}
\newcommand{\ee}{\end{equation}}
\newcommand{\ba}{\begin{array}{c}}
\newcommand{\ea}{\end{array}}
\newcommand{\bqa}{\begin{eqnarray}}
\newcommand{\eqa}{\end{eqnarray}}

\makeatletter
    
    \newcommand{\Rmnum}[1]{\expandafter\@slowromancap\romannumeral #1@}
    \makeatother

\begin{document}

\bibliographystyle{unsrt}

\title{\bf The $1/N_c$ expansion in hadron effective field theory }

\author{Guo-Ying Chen}

\affiliation{Department of Physics and Astronomy, Hubei University
of Education, Wuhan 430205, China}

\email{chengy@pku.edu.cn}

\date{\today}

\begin{abstract}
We study the $N_c$ scalings of pion-nucleon and nucleon-nucleon
scatterings in hadron effective field theory. By assuming Witten's
counting rules are applied to matrix elements or scattering
amplitudes which use the relativistic normalization for the
nucleons, we find that the nucleon axial coupling $g_A$ is of order
$N_c^0$, and a consistent large $N_c$ counting can be established
for the pion-nucleon and nucleon-nucleon scatterings. We also
justify the nonperturbative treatment of the low energy
nucleon-nucleon interaction with the large $N_c$ analysis and find
that the deuteron binding energy is of order $1/N_c$.
\end{abstract}


\maketitle

In the seminal paper Ref.~\cite{'tHooft:1973jz}, 't Hooft showed
that QCD has a hidden expansion parameter $1/N_c$. With the $1/N_c$
expansion the QCD coupling constant $g$ is of order $1/\sqrt{N_c}$,
and gluons are denoted by double lines. By counting the number of
interaction vertexes and closed color loops, one can figure out the
$N_c$ order of a specific Feynman diagram. It is then found that in
the large $N_c$ limit, the leading contribution comes from planar
diagrams with minimal quark loops. The idea to expand QCD in $1/N_c$
is very attractive, as it explains hadron phenomenology
successfully, for instance the OZI suppression rule. The extension
of the $1/N_c$ expansion to baryons was first carried out by
Witten~\cite{Witten:1979kh}. The $1/N_c$ expansion of baryons, which
are bound states of $N_c$ valence quarks and have masses of order
$N_c$, is more complicated than that of mesons. Using quarks and
gluons as degrees of freedom, Witten showed that meson-baryon
coupling is of order $\sqrt{N_c}$, meson-baryon scattering amplitude
is of order $N_c^0$, and baryon-baryon scattering amplitude is of
order $N_c$~(these results are called the large $N_c$ counting rules
of Witten in this manuscript). It is interesting to study whether a
realistic hadron effective field theory using baryons and mesons as
degrees of freedom can reproduce the large $N_c$ counting rules of
Witten. We will see in the following that this is not a clearly
solved problem.

\begin{figure}[hbt]
\begin{center}
  \includegraphics[width=10cm]{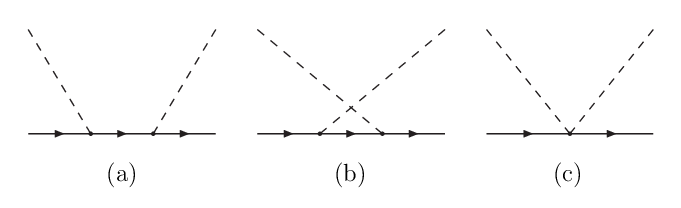}\\
  \caption{Pion-nucleon scattering diagrams in leading chiral expansion.}\label{piN}
  \end{center}
\end{figure}

First of all, we consider the pion-nucleon scattering. The leading
pion-nucleon scattering diagrams in chiral expansion are shown in
Fig.~\ref{piN}. In chiral theory, the pion-nucleon vertex is
proportional to the factor $\frac{g_A}{f_\pi}$, where $f_\pi$ is the
pion decay constant with order $\sqrt{N_c}$, and $g_A$ is the
nucleon axial coupling constant which is assumed to be of order
$N_c$ in both the nonrelativistic quark model\cite{Manohar:1998xv}
and the skyrme model~\cite{Adkins:1983ya}. The nucleon propagator is
of order $N_c^0$~\cite{Jenkins:1998wy}. Therefore, both amplitudes
of Fig.~\ref{piN}(a) and (b) are proportional to the factor
$(\frac{g_A}{f_\pi})^2$ and are of order $N_c$, and
Fig.~\ref{piN}(c) is proportional to $1/f_\pi^2$ and of order
$1/N_c$. Clearly these results contradict the large $N_c$ counting
rules of Witten. An solution to this puzzle has been proposed in
Ref.~\cite{Dashen:1993as,Gervais:1983wq,Gervais:1984rc}, and we give
a short review here for the convenience of further discussions. The
axial vector current matrix element in the nucleon is of order $N_c$
and can be written as~\cite{Dashen:1993as}
\begin{equation}
\langle N|\bar \psi \gamma^i\gamma_5\tau^a \psi|N\rangle=N_c g
\langle N|X^{ia}|N\rangle,\label{ACME}
\end{equation}
where the $N_c$ dependence has been explicitly factored out, so
$\langle N|X^{ia}|N\rangle$ and $g$ are of order one. The sum of the
scattering amplitudes for the pole diagrams of
$\pi^a(q)+N(P)\rightarrow \pi^b(q^\prime)+N(P^\prime)$ reads
\begin{equation}
-iq^iq^{\prime j}\frac{N_c^2
g^2}{f_\pi^2}\left[\frac{1}{q^0}X^{jb}X^{ia}-\frac{1}{q^{\prime
0}}X^{ia}X^{jb}\right],\label{poleAMP}
\end{equation}
where the amplitude is written in the form of an operator acting on
nucleon states, and $q^0=q^{\prime 0}$ as both initial and final
nucleons are on-shell. One can see explicitly that the above
amplitude is of order $N_c$ and violates the unitarity. The solution
to this puzzle is the commutation relation in the large $N_c$ limit
\begin{equation}
[X^{ia},X^{jb}]=0,\label{LNcCC}
\end{equation}
which is usually called the large $N_c$ consistency condition. The
solution of the large $N_c$ consistency condition requires to
consider all the possible baryon states degenerate with the nucleon
in the intermediate states of the scattering. For instance, besides
the nucleon, $\Delta$ should also be included as an intermediate
state in the pion-nucleon scattering, and the coupling $g_{\pi
N\Delta }$ can be determined in terms of $g_{\pi NN}$ using the
large $N_c$ consistency condition. The large $N_c$ consistency
condition can be derived from the spin-flavor algebra. Consider the
two-flavor case, the $SU(4)$ generators $J^i,I^a$ and $G^{ia}$
satisfy the spin-flavor algebra
\begin{eqnarray}
\left[J^i,J^j\right]&=&i\epsilon^{ijk}J^k,\ \
\left[I^a,I^b\right]=i\epsilon^{abc}I^c,\ \ \left[I^a,G^{ib}\right]=i\epsilon^{abc}G^{ic},\nonumber\\
\left[J^i,G^{ja}\right]&=&i\epsilon^{ijk}G^{ka},\ \
\left[J^i,I^a\right]=0,\ \
\left[G^{ia},G^{jb}\right]=\frac{i}{4}\delta^{ab}\epsilon^{ijk}J^{k}+\frac{i}{4}\delta^{ij}\epsilon^{abc}I^c.\label{SFA}
\end{eqnarray}
Rescaling the generator $G^{ia}$ by a factor of $1/N_c$ and taking
the large $N_c$ limit,
\begin{equation}
X^{ia}=\lim_{N_c\rightarrow \infty}\frac{G^{ia}}{N_c},
\end{equation}
we can find that the commutation relation for $[G^{ia},G^{jb}]$
turns into Eq.~(\ref{LNcCC}) in the large $N_c$ limit, as the order
of baryon matrix elements of $J$ and $I$ are at most of $N_c$. A
careful study find that the commutator $[X^{ia},X^{jb}]$ is of order
$1/N_c^2$~\cite{Dashen:1993ac}, hence Eq.~(\ref{poleAMP}) is of
order $1/N_c$, which is the same as the amplitude of
Fig.~\ref{piN}(c). Since Witten's counting rules suggest that the
leading contribution should come from order $N_c^0$ diagrams in the
$1/N_c$ expansion, one may wonder why the pion-nucleon scattering
amplitude is dominated by order $1/N_c$ diagrams instead of order
$N_c^0$ diagrams.

\begin{figure}[hbt]
\begin{center}
  \includegraphics[width=7cm]{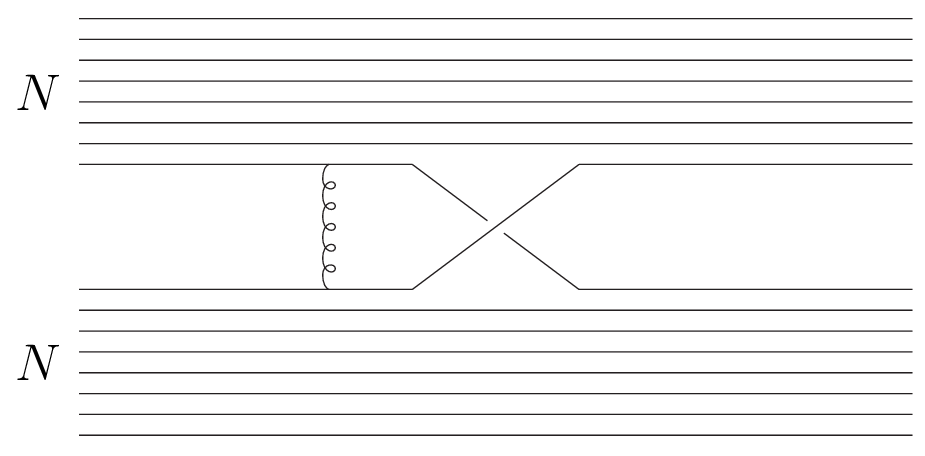}\\
  \caption{A typical diagram of order $N_c$ which contributes to the nucleon-nucleon interaction in large $N_c$
  QCD.}\label{BB}
  \end{center}
\end{figure}

Secondly, there is difficulty in matching the $N_c$ counting of the
nucleon-nucleon potential calculated at the quark-gluon level to
that calculated at the hadronic level. The dominant nucleon-nucleon
interaction is generically of order $N_c$ with the analysis at the
quark-gluon level~\cite{Witten:1979kh}. This can be understood from
the one-gluon exchange diagrams in Fig.~\ref{BB}, where each gluon
coupling has a factor $1/\sqrt{N_c}$, and there are $N_c^2$ ways to
choose a pair of quarks. In studying the $N_c$ counting of the
nucleon-nucleon potential, Ref.~\cite{Kaplan:1995yg,Kaplan:1996rk}
consider the kinematic region ${\bf{p}}\sim N_c^0$ for the nucleon
and assume that quark-line connected diagrams such as Fig.~\ref{BB}
are of order $N_c$ leads to a potential of order $N_c$. It should be
note that the kinematic region is different from that was considered
in Witten's paper, where ${\bf{p}}\sim N_c$~\cite{Witten:1979kh}.
The reason is that, if ${\bf{p}}\sim N_c$ which is the same order as
the nucleon mass, the nucleon is a relativistic particle, and this
kinematic region is beyond the scope of the low energy hadron
effective field theory~\cite{Banerjee:2001js}. At the hadronic
level, it seems straightforward to see that the one-meson exchange
potential is of order $N_c$, for example the one-pion exchange
potential is proportional to the factor $(\frac{g_A}{f_\pi})^2$ and
of order $N_c$. At the two-meson exchange level, diagrams(for
instance, the crossed-box diagram) with four nucleon-meson vertexes
are of order $N_c^2$ and will contribute to the potential, but
Ref.~\cite{Banerjee:2001js} shows that there is a cancelation of the
retardation effect of the box graph against the contribution of the
crossed-box diagram, therefore the dominant nucleon-nucleon
potential remains to be of order $N_c$. Nevertheless, it is
difficult to show that whether such cancelations will happen in all
multi-meson exchange levels. Actually, it is found that at the
three- and higher-meson exchange level the potential derived from
the hadronic theory can be larger than that of the quark-gluon
theory~\cite{Belitsky:2002ni}. This problem is called the large
$N_c$ nuclear potential puzzle.

Two possible resolutions to the large $N_c$ nuclear potential puzzle
are suggested in Ref.~\cite{Belitsky:2002ni}. One possibility is
that necessary cancelations might happen if the hadronic-level
calculation is reorganized in some other way. A tentative study on
this direction is done in Ref.~\cite{Cohen:2002im}, where the
hadronic-level calculation is reorganized in the way that the
potential is energy-independent. The other possibility is that the
$N_c$ scaling rule of the nucleon-nucleon potential proposed in
Ref.~\cite{Kaplan:1995yg,Kaplan:1996rk} may be invalid. This can be
plausible by noting that some nuclear phenomena may be difficult to
understand if the nucleon potential is of order $N_c$, for example,
the deuteron binding energy is a small number in reality, i.e.,
$B=2.2$ MeV. Although it is generally believed that such small
binding energy results from delicate cancellation which occurs at
$N_c=3$~\cite{Gross:2011ve,Cohen:2013tya}, it will still be
interesting to find an alternative solution which does not rely on
the fact that $N_c=3$ in the real world. In addition, since the
nucleon kinetic energy is order $N_c^{-1}$, if the potential is
order $N_c$ nucleon matter forms a crystal in the large $N_c$ limit.
However, nuclear matter appears to be in liquid state rather than in
crystal state in reality. Generally, nuclear phenomena seem to
prefer a small potential value, and this observation motivates
Ref.~\cite{Hidaka:2010ph,Kojo:2012hf} to proposed a refined quark
model which gives $g_A\sim N_c^0$, thus the one-pion exchange
potential is of order $N_c^{-1}$.

From the above discussions, we can see that it is not clear how to
reproduce the large $N_c$ counting rules of Witten consistently in a
realistic hadron effective field theory. In this paper we try to
propose a simple way to resolve the above difficulties. We find that
we can have a consistent large $N_c$ counting in hadron effective
field theory if we assume Witten's counting rules are applied to
matrix elements or scattering amplitudes which use the relativistic
normalization for the nucleons. Our work is organized as following:
we first discuss the difference of $N_c$ scalings between the
relativistic and nonrelativistic normalizations. With the
relativistic normalization, we then discuss the $N_c$ scalings of
pion-nucleon and nucleon-nucleon scatterings. We finally extend our
study to the meson-meson scatterings and find that loosely-bound
meson-meson molecular states may not exist in the large $N_c$ limit.

The leading non-relativistic chiral Lagrangian for the pion-nucleon
coupling reads
\begin{equation}
\mathcal{L}_{\pi
N}=\frac{g_A}{2f_{\pi}}\psi^\dagger\vec{\sigma}\cdot(\vec{\partial}\pi)\psi,
\end{equation}
where $\psi$ is the nucleon doublet and $\pi=\tau^i\pi^i$. If $g_A$
is of order $N_c$, the pion-nucleon coupling is proportional to
$\frac{g_A}{f_\pi}$ and of order $\sqrt{N_c}$, which is just the
result of Witten. But if $g_A\sim N_c^0$ as suggested in
Ref.~\cite{Hidaka:2010ph,Kojo:2012hf}, the pion-nucleon coupling is
of order $N_c^{-1/2}$, which seems to contradict the large $N_c$
counting rules of Witten. One should note that in the above $N_c$
counting Feynman rules for the external nucleon lines are assumed to
be independent of $N_c$, or explicitly one use the nonrelativistic
normalization for the nucleon
\begin{equation}
\langle
p(\vec{k},s_2)|p(\vec{p},s_1)\rangle=\delta^3(\vec{k}-\vec{p})\delta_{s_1,s_2}.\label{nm}
\end{equation}
However from the Dirac theory, we know that the Feynman rule for the
external nucleon line is $u_s(p)$, which reduces to
$\sqrt{2m_N}\chi_s$($\chi_s$ is the two-component spinor) in the
nonrelativistic limit and obviously depends on $N_c$. Thus the
relativistic normalization for the nucleon reads
\begin{equation}
\langle p(\vec{k},s_2)|p(\vec{p},s_1)\rangle=
2m_N\delta^3(\vec{k}-\vec{p})\delta_{s_1,s_2},\label{normalize}
\end{equation}
where we have taken the nonrelativistic limit for the Dirac spinor
$u_s(p)$ to simplify the $N_c$ counting in the low energy effective
field theory. Scattering amplitudes which use Eq.~(\ref{normalize})
as the normalization condition can be understood as the
nonrelativistic reduction of the relativistic form. In this work, we
will use the relativistic normalization for the scattering
amplitudes, thus the relation between our defined scattering
amplitude $\mathcal{M}(p_1,p_2\rightarrow p_f)$ and the $S$-matrix
reads
\begin{eqnarray}
S=1+i\mathcal{T},
 \end{eqnarray}
where
\begin{equation}
<p_f|i\mathcal{T}|p_1,p_2>=i(2\pi)^4\delta^{(4)}(p_1+p_2-\sum
p_f)\mathcal{M}(p_1,p_2\rightarrow p_f).
\end{equation}
To obtain the scattering amplitudes with the nonrelativistic
normalization $\mathcal{M}_{NR}$, one takes the limit
$|\vec{p}|\rightarrow 0$ and drops factors of $\sqrt{2m_N}$ which
come from external nucleon spinor function $u_s(p)$. For example,
for the nonrelativistic nucleon-nucleon elastic scattering, we can
have $\mathcal{M}_{NR}=\mathcal{M}/(2m_N)^2$. One should note that
we only discuss the nonrelativistic scattering in this work as the
relativistic normalization reduce to the form in
Eq.(\ref{normalize}). We will find that Witten's large $N_c$
counting rules can be consistent, if one assume that these rules are
applied to matrix elements or scattering amplitudes which use the
relativistic normalization for the nucleon.

We now return back to the discussion of the pion-nucleon coupling.
With the normalization in Eq.(\ref{normalize}), there is a factor of
$\sqrt{2m_N}$ for each external nucleon line, and the pion-nucleon
coupling is proportional to $m_N\frac{g_A}{f_\pi}$. Our matching
scheme is that we assume Witten's counting rules are applied to
matrix elements or scattering amplitudes which use the relativistic
normalization for the nucleon, we then find that $g_A\sim N_c^0$ as
the pion-nucleon coupling is order $\sqrt{N_c}$ from Witten's rules.
Thus we have shown that if $g_A\sim N_c^0$ as suggested in
Ref.~\cite{Hidaka:2010ph,Kojo:2012hf}, the $N_c$ counting of the
pion-nucleon coupling can be consistent with Witten's result. In the
following we will count $g_A$ as order $N_c^0$, since it is the
consequence of our assumption.

Now we come to discuss the $N_c$ scaling of the pion-nucleon
scattering. The scattering amplitude for the pion-nucleon scattering
given in Eq.~(\ref{poleAMP}) is written with the nonrelativistic
normalization. But noting that, while factors of $\sqrt{2m_N}$ which
come from the external nucleon lines are dropped in
Eq.~(\ref{poleAMP}), such factor is kept for the baryon propagator.
This can be obviously since the baryon propagator $\frac{i}{q^0}$ in
Eq.~(\ref{poleAMP}) is obtained from the Dirac propagator by taking
the nonrelativistic limit
\begin{eqnarray}
\frac{i(\slashed{P}+\slashed{q}+m_B)}{(P+q)^2-m_B^2}\rightarrow
\frac{i} {q\cdot v}(\frac{1+\slashed{v}}{2}),
\end{eqnarray}
where $m_B$ is the mass of intermediate baryon, $v^\mu$ is the
four-velocity of the initial nucleon, i.e., $P^\mu=m_N v^\mu$, and
it is taken to be $v^\mu=(1,\vec{0})$ in the nonrelativistic limit.
With the the relativistic normalization, factors of $\sqrt{2m_N}$
should be restored for external nucleon lines in the scattering
amplitudes, then the scattering amplitudes for Fig.~\ref{piN}(a,b)
are proportional to $m_N(\frac{g_A}{f\pi})^2$, and the amplitude for
Fig.~\ref{piN}(c) is proportional to $m_N/f_\pi^2$, all of which are
of order $N_c^0$ and consistent with the large $N_c$ counting rules
of Witten. Actually what we have illustrated in the above is that if
the assumption that Witten's rules are applied to matrix elements or
scattering amplitudes which use the relativist normalization for the
nucleon is adopted for the pion-nucleon coupling, the $N_c$ scaling
of the pion-nucleon scattering will be consistent with this
assumption.

To have a consistent $N_c$ counting with the nonrelativistic
normalization, one can redefine the nucleon field by dividing a
factor of $\sqrt{2m_N}$, then factors of $\sqrt{2m_N}$ are not
needed in the Feynman rules for the external nucleon lines.
Meanwhile, the nucleon propagator should be $\frac{i}{2m_N q^0}$
instead of $\frac{i}{q^0}$ as the new defined nucleon field has the
dimension one, and $g_A$ should absorb a factor of $2M_N$ which is
dimension one and of order $N_c$. With this convention the
scattering amplitudes for Fig.~\ref{piN}(a,b) are proportional to
$(\frac{g_A}{f\pi})^2\frac{1}{2m_N q^0}$, which is also of order
$N_c^0$ as $g_A$ is of the order $N_c$. Therefore, we can see that
physical results do not depend on the normalization conventions as
long as they are used consistently. The advantage of using the
relativistic normalization is that $g_A$ is a dimensionless constant
as usually defined in the text book.

It is worth to mention that, although with the relativistic
normalization $g_A$ is of order $N_c^0$, the commutation relation in
Eq.~(\ref{LNcCC}) still holds in the large $N_c$ limit. This can be
shown by rewriting the axial current matrix element in
Eq.~(\ref{ACME}) with the normalization in Eq.~(\ref{normalize})
\begin{equation}
\langle N|\bar \psi \gamma^i\gamma_5\tau^a \psi|N\rangle=2m_N g
\langle N|X^{ia}|N\rangle=N_c\frac{2m_N}{N_c}g\langle
N|X^{ia}|N\rangle,\label{NACME}
\end{equation}
where $\langle N|X^{ia}|N\rangle$ and $g$ are of order one, and the
product $g\langle N|X^{ia}|N\rangle$ is different from that in
Eq.~(\ref{ACME}) by a $N_c$ independent factor $\frac{2m_N}{N_c}$.
Since the  matrix element of $X^{ia}$ defined in Eq.~(\ref{NACME})
is different from that in Eq.~(\ref{ACME}) by a $N_c$ independent
factor, operators $X^{ia}$ defined in Eq.~(\ref{NACME}) still
satisfy the commutation relation in Eq.~(\ref{LNcCC}), i.e., the
large $N_c$ consistency condition. The commutation relation
Eq.~(\ref{LNcCC}) arises because the axial vector current matrix
element grows with $N_c$. With our normalization, $g_A$ is of order
$N_c^0$, but the axial vector current matrix element is still of
order $N_c$, thus the commutation relation Eq.~(\ref{LNcCC}) still
holds. Taking this contracted spin-flavor symmetry into account, we
should also include $\Delta$ particle in the intermediate state,
then the contributions from pole diagrams Fig.~\ref{piN}(a) and (b)
vanish in the large $N_c$ limit.

\begin{figure}[hbt]
\begin{center}
  \includegraphics[width=10cm]{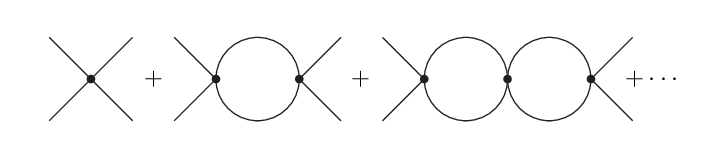}\\
  \caption{Feynman diagrams generated by the contact interaction.}\label{NNscattering}
\end{center}
\end{figure}

We then come to the nucleon-nucleon scatterings. As in the above, we
assume Witten's counting rules are applied to the nucleon-nucleon
scattering amplitudes which use the relativistic normalization. In
this way, nucleon-nucleon scattering amplitudes with the
relativistic normalization are of order $N_c$. The order $N_c$
scattering amplitude leads to a potential of order $N_c^{-1}$, as
factors of $\sqrt{2m_N}$ which come from the external nucleon lines
should be dropped in the scattering amplitude to obtain the
non-relativistic potential. Therefore our resolution to the nuclear
potential puzzle corresponds to the second possibility suggested in
Ref.~\cite{Belitsky:2002ni}, i.e., the assumption that the dominant
nucleon-nucleon potential is of order $N_c$ is somewhat heuristic
and may be invalid. We will take the pion-exchange as an example to
show explicitly how the large $N_c$ nuclear potential puzzle can be
resolved. The nucleon-nucleon scattering amplitude for the one-pion
exchange diagram is proportional to the factor
$(m_N\frac{g_A}{f_\pi})^2$, which is of order $N_c$ and again
consistent with our assumption. The amplitude of two-pion exchange
crossed-box diagram is proportional to $m_N^2(\frac{g_A}{f_\pi})^4$,
which is of order $N_c^0$ and thus is suppressed by $1/N_c$. It is
obvious that more-pion exchange diagrams which are two-nucleon
inreducible, will be suppressed by more powers of $1/N_c$. Thus one
can see that the nucleon-nucleon potential is at most of order
$N_c^{-1}$, and this large $N_c$ counting rule will not be violated
at the multi-pion exchange level. It is straightforward to see that
similar conclusion can be obtained for other meson-exchange
diagrams, in particularly the sigma-exchange potential is also of
order $N_c^{-1}$.

So far, we have shown that a consistent large $N_c$ counting in
hadron effective field theory can be established, if we assume
Witten's counting rules are applied to matrix elements or scattering
amplitudes which use the relativistic normalization for the
nucleons. Nevertheless there are still two points in nucleon-nucleon
scatterings need to be further investigated. The first point is that
one may wonder whether the order $N_c$ scattering amplitude violates
the unitarity. We now show that this does not happen in the
nucleon-nucleon scattering. To be specific, let's consider the
single channel $S$-wave nucleon-nucleon scattering. The $S$-matrix
for the $S$-wave nucleon-nucleon scattering can be written as
$S=1+i\frac{p}{8\pi m_N}\mathcal{M}$, where $p$ is the nucleon
momentum and $\mathcal{M}$ is the scattering amplitude which use the
relativistic normalization. Unitary condition, i.e. $SS^\dagger=1$,
requires that
\begin{equation}
\mbox{Im}\mathcal{M}=\frac{p}{16\pi m_N}|\mathcal{M}|^2.
\end{equation}
We consider the kinematic region $p\sim N_c^0$ as in
Ref.~\cite{Kaplan:1996rk} and denote the $N_c$ scaling of the
amplitude as $\mbox{Re}\mathcal{M}\sim
N_c^{n_1},\mbox{Im}\mathcal{M}\sim N_c^{n_2}$. The unitary condition
in the large $N_c$ limit can then be written as
\begin{equation}
n_2=2\mbox{max}(n_1,n_2)-1.
\end{equation}
The solution to the above condition reads
\begin{eqnarray}
\mbox{If}\ \ \  n_2&\geqslant& n_1,\ \ \  \mbox{then}\ \ \
n_1\leqslant n_2=1,\
\ \ \mathcal{M}\sim \mathcal{O}(N_c);\nonumber\\
\mbox{If}\ \ \ n_2&<&n_1,\ \ \ \mbox{then}\ \ \ n_2<n_1<1,\ \ \
\mathcal{M}<\mathcal{O}(N_c).
\end{eqnarray}
Thus one can see that the unitary condition can be satisfied if the
scattering amplitude is of order $N_c$.

The second point is that in hadron effective field theory deuteron
corresponds to a bound state pole which emerges from the
non-perturbative summation of the nucleon-nucleon scattering
amplitudes, hence it is interesting to investigate whether such
non-perturbative summation is justified in the large $N_c$ limit.
Following, we will study this point in detail with a low energy
pionless effective field theory.

\begin{figure}[hbt]
\begin{center}
  \includegraphics[width=8cm]{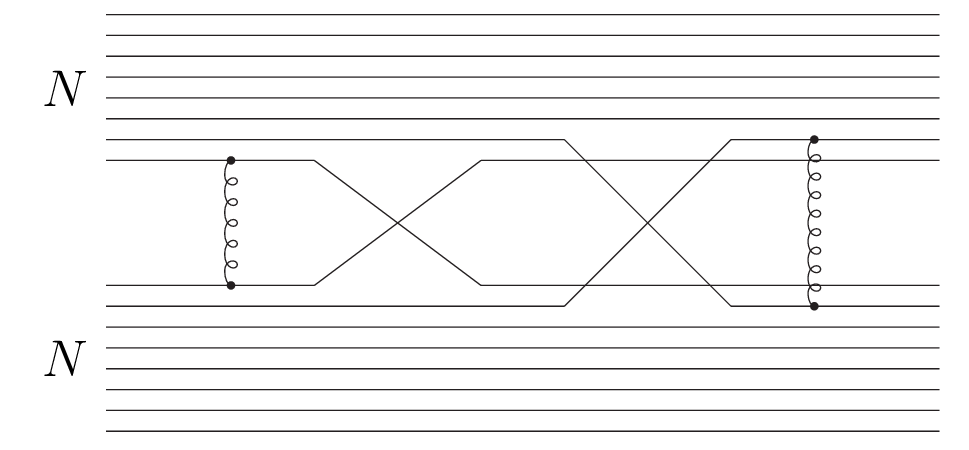}\\
  \caption{A typical quark-disconnected diagram.}\label{quarkDC}
  \end{center}
\end{figure}

Let's now consider the low energy pionless effective field theory in
the $^3S_1$ channel~\cite{Wein,KSW,vanKolck:1999mw}. The leading
order operator in the chiral expansion for nucleon-nucleon
interactions reads
\begin{equation}
\mathcal{L}_{NN}=-C_0(\psi^\dagger \psi)^2.
\end{equation}
Feynman diagrams generated by this interaction vertex are shown in
Fig.~\ref{NNscattering}. We can identify the order $N_c$
quark-connected diagrams, such as Fig.~\ref{BB}, as the tree diagram
in Fig.~\ref{NNscattering}, and identify the quark-disconnected
diagrams, such as Fig.~\ref{quarkDC}, as the one-loop diagram in
Fig.~\ref{NNscattering}. The amplitude for the tree diagram in
Fig.~\ref{NNscattering} reads
\begin{equation}
i\mathcal{M}_{tree}=-i4m_N^2 C_0,
\end{equation}
where the factor $4m_N^2$ comes from four external nucleon lines.
$\mathcal{M}_{tree}$ is of order $N_c$, according to our assumption
that Witten's counting rules are applied to the scattering
amplitudes which use the relativistic normalization. We can then
find that $C_0$ is of order $N_c^{-1}$. Amplitudes for the loop
diagrams can be obtained after treating the loop integral. In the
pionless effective field theory, the loop integral in can be done in
the nonrelativistic approximation. Using the minimal subtraction
scheme, the loop integral reads~\cite{KSW}
\begin{eqnarray}
\mathcal{I}&=&\int\frac{d^D\ell}{(2\pi)^D}\frac{i}{\ell^0-\vec{\ell}^2/(2m_N)+i\epsilon}\cdot
\frac{i}{E_k-\ell^0-\vec{\ell}^2/(2m_N)+i\epsilon},\nonumber\\
&=&\int\frac{d^{D-1}\ell}{(2\pi)^{D-1}}\frac{i}{E_k-\vec{\ell}^2/(m_N)+i\epsilon},\nonumber\\
&=&i\frac{m_N}{4\pi}(-m_N E_k-i\epsilon)^{1/2},\nonumber\\
&=&\frac{m_N p}{4\pi},
\end{eqnarray}
where $E_k=\vec{p}^2/m_N$ is the total kinetic energy of the
two-nucleon in the center of mass frame. In the chiral expansion the
three-momentum of the nucleon is treated as small scale and has the
same order as $m_\pi$, thus we assume $p$ to be independent of $N_c$
as in Ref.~\cite{Kaplan:1996rk}. It is worth mentioning that the
nucleon propagator used here is order $N_c$ and different from that
in Eq.~(\ref{poleAMP}), as the term $\vec{\ell}^2/2m_N$ should also
be included in the propagator to avoid the infrared divergence in
the two-nucleon reducible diagrams~\cite{Wein}.  With the result for
loop integral, one can obtain the one-loop amplitude
\begin{equation}
i\mathcal{M}_{loop}=-\frac{p}{16\pi
m_N}\mathcal{M}_{tree}^2=-\frac{1}{\pi}m_N^3 C_0^2p.
\end{equation}
We can find that the one-loop amplitude is of order $N_c$, which is
the same as that of the tree diagram. This result is somewhat
surprising at the first sight, as quark-disconnected diagram
Fig.~\ref{quarkDC} seems to be of order $N_c^2$. Actually to count
the $N_c$ order of Fig.~\ref{quarkDC} one should note that such a
diagram contains nucleon propagators and loop, thus its $N_c$ order
can be the same as that obtained in the hadron effective field
theory, i.e., $\mathcal{O}(N_c)$. It is straightforward to count the
$N_c$ order of all the other diagrams in Fig.~\ref{NNscattering}.
One can find that all the diagrams in Fig.~\ref{NNscattering} are of
order $N_c$, hence the amplitude at the leading order in the $1/N_c$
expansion should come from the non-perturbative summation of all the
diagrams in Fig.~\ref{NNscattering}. The re-summed amplitude reads
\begin{equation}
i\mathcal{M}_{sum}=-i\frac{4m_N^2 C_0}{1+i\frac{C_0 m_N
}{4\pi}p},\label{resumAMP}
\end{equation}
which is obviously of order $N_c$. The partial wave $S$-matrix can
then be written as
\begin{equation}
S=1+i\frac{p}{8\pi m_N}\mathcal{M}_{sum}=1-i\frac{m_N
p}{2\pi}\frac{C_0}{1+i\frac{C_0 m_N}{4\pi}p}.
\end{equation}
One can find that, although $\mathcal{M}_{sum}$ is of order $N_c$,
the S-matrix is independent of $N_c$ and satisfies the unitary
condition, i.e. $SS^\dagger=1$. Actually it has already been shown
in Ref.~\cite{KSW} that all the diagrams in Fig.~\ref{NNscattering}
need to be summed, because they all are at the leading chiral order
$\mathcal{O}(p^{-1})$. Here we have shown that this summation is
also justified in the large $N_c$ expansion. The deuteron
corresponds to a bound state pole at $E_k=-B$ in
Eq.(\ref{resumAMP}), where $B$ is the binding energy,
\begin{equation}
B=\frac{16\pi^2}{C_0^2 m_N^3},\label{Benergy}
\end{equation}
which is of order $N_c^{-1}$. Similar analysis has been done in
Ref.~\cite{Beane:2002ab}, but the conclusion is different. Because
$C_0$ is taken to be $\mathcal{O}(N_c)$, $B$ is found to be
$\mathcal{O}(N_c^{-5})$ in Ref.~\cite{Beane:2002ab}. However, if
$C_0$ is order $N_c$ and $B$ is order $N_c^{-5}$, it will then be a
puzzle why the binding energy is so small, while the nonrelativistic
potential is large. We can also study the $N_c$ scaling of the
deuteron binding energy in the pionfull effective field theory with
the method used in Ref.~\cite{Beane:2002ab}. In the chiral limit,
the Schr\"{o}dinger equation in coordinate space can be simply
written as~\cite{Beane:2002ab}
\begin{equation}
(\frac{\nabla^2}{m_N}-\frac{3\alpha_\pi}{r^3})|\Psi>=-B|\Psi>,\label{SchEq}
\end{equation}
where
\begin{equation}
\alpha_\pi=\frac{g_A^2}{16\pi f_\pi^2}.
\end{equation}
Because the three-momentum of the nucleon and $g_A$ are of order
$N_c^0$, both kinetic energy and tensor potential energy in the left
hand side of Eq.(\ref{SchEq}) are of order $N_c^{-1}$, and the
binding energy $B$ is of order $N_c^{-1}$ which is the same as that
in the pionless effective field theory. This conclusion remains
unchanged even if an additional term $C_0\delta^{(3)}(r)$ is
included in the potential. In contrast, by treating $g_A$ as
$\mathcal{O}(N_c)$, Ref.~\cite{Beane:2002ab} assumes that the
coordinate scales as $N_c^2$, or equivalently the three-momentum
carried by the nucleon scales as $1/N_c^2$, then $B$ scales as
$N_c^{-5}$. However, as mentioned in Ref.~\cite{Beane:2002ab}, if
the three-momentum carried by the nucleon scales as $1/N_c^2$, the
effective field theory will be constrained to threshold.

It is interesting to extend the above analysis to discuss the
existence of $S$-wave meson-meson molecular states. We will find
that the $N_c$ analysis of the deuteron cannot be straightforwardly
extended to the meson-meson case as the meson-meson scattering
amplitudes and meson masses have different $N_c$ countings. In
effective field theory approach, an $S$-wave meson-meson molecular
state corresponds to a bound state pole in the elastic meson-meson
scattering amplitude coming from the summation of all the diagrams
in Fig.~\ref{NNscattering}~\cite{AlFiky:2005jd}. Similar to
Eq.~(\ref{Benergy}), the binding energy $\tilde{B}$ for the
meson-meson molecular state reads
\begin{equation}
\tilde{B}=\frac{2\pi^2}{\tilde{C}_0^2\mu^3},
\end{equation}
where $\tilde{C}_0$ is the coefficient of the contact four-meson
operator, and $\mu$ is the reduced mass of the two-meson system.
$\tilde{C}_0$ is of order $N_c^{-1}$, as the tree level amplitude
for the meson-meson scattering scales as $N_c^{-1}$. $\mu$ is of
order $N_c^0$, as the meson mass is independent of $N_c$. Therefore,
we can find that the binding energy of the $S$-wave meson-meson
molecular state scales as $N_c^2$, and the binding momentum is of
order $N_c$. This indicates that if the meson-meson molecular state
exists, it should be deeply bounded. However, mesons with the
momenta of order $N_c$ are extremely-relativistic particles, and
their scatterings cannot be treated in the nonrelativistic effective
field theory. If the momentum of the meson is taken to be
independent of $N_c$ as that of the meson mass, we will find that
the diagrams in Fig.~\ref{NNscattering} have different $N_c$
scalings. The tree diagram in Fig.~\ref{NNscattering} scales as
$N_c^{-1}$, and the $n$-loop diagram are suppressed which scales as
$N_c^{-(n+1)}$. These results are different from the $S$-wave
nucleon-nucleon scattering, due to the fact that the meson mass is
of order $N_c^0$, while the nucleon mass is of order $N_c$. Thus for
meson-meson scatterings, only the tree level diagram contributes to
the leading order amplitude in the $1/N_c$ expansion, which is just
the conclusion in the standard large $N_c$ analysis for meson-meson
interactions~\cite{Witten:1979kh}, and the summation of all the
diagrams in Fig.~\ref{NNscattering} is unnecessary. We then conclude
that, there is no loosely-bound meson-meson molecular state in the
large $N_c$ limit, as the meson-meson interaction is weak, and
similar conclusion has also been given in Ref.~\cite{Liu:2007tj}.
Finally, we would like to mention that we only discuss the existence
of the meson-meson molecular state in the large $N_c$ limit, for
other configurations such as tetraquark and polyquark states one can
refer to Ref.~\cite{Cohen:2014vta} and references therein.

In summary, we have tried to propose a possible solution to overcome
the difficulties which are encountered in reproducing the large
$N_c$ counting rules of Witten in hadron effective field theory. We
find that a consistent large $N_c$ counting can be established if we
assume Witten's counting rules are applied to matrix elements or
scattering amplitudes which use the relativistic normalization for
the nucleons. We also find that at the leading order in the $1/N_c$
expansion, the $S$-wave nucleon-nucleon scattering should be treated
nonperturbatively, and the deuteron binding energy is of order
$1/N_c$ which is consistent with the nuclear phenomena. In contrast,
the $S$-wave meson-meson interaction is weak, and loosely-bound
meson-meson molecular states may not exist in the large $N_c$ limit.


\section*{ACKNOWLEDGMENTS}
I would like to thank Yu Jia for helpful discussions and J.Ruiz de
Elvira for valuable comments. Part of this work was done during my
visit to Institute of Theoretical Physics, Chinese Academy of
Science in Beijing. This work is supported, in part, by National
Natural Science Foundation of China (Grant Nos. 11305137).

\end{document}